\def\spose#1{\hbox to 0pt{#1\hss}}
\def\approxlt{\mathrel{\spose{\lower 3pt\hbox{$\sim$}}
        \raise 2.0pt\hbox{$<$}}}
\def\approxgt{\mathrel{\spose{\lower 3pt\hbox{$\sim$}}
        \raise 2.0pt\hbox{$>$}}}

\def\mdot{\hbox{$\dot m$}}
\def\cm{{\rm\thinspace cm}}

\def\erg{{\rm\thinspace erg}}

\def\keV{{\rm\thinspace keV}}
\def\km{{\rm\thinspace km}}

\def\Mpc{{\rm\thinspace Mpc}}
\def\Msun{\hbox{$\rm\thinspace M_{\odot}$}}

\def\s{{\rm\thinspace s}}

\def\sr{{\rm\thinspace sr}}

\def\ergpcmsqps{\hbox{$\erg\cm^{-2}\s^{-1}\,$}}

\def\ergps{\hbox{$\erg\s^{-1}\,$}}

\def\kmps{\hbox{$\km\s^{-1}\,$}}

\def\pcmsq{\hbox{$\cm^{-2}\,$}}

\def\pmpccu{\hbox{$\Mpc^{-3}\,$}}
\def\ps{\hbox{$\s^{-1}\,$}}

\def\psr{\hbox{$\sr^{-1}\,$}}

\def\kmpspMpc{\hbox{$\kmps\Mpc^{-1}$}}

\documentclass[]{article}
\usepackage{emulateapj}
\usepackage{psfig}

\begin{document}

\title{HARD X-RAY EMISSION FROM ELLIPTICAL GALAXIES AND ITS CONTRIBUTION 
TO THE X-RAY BACKGROUND}

\author{Tiziana Di Matteo\altaffilmark{1}}

\affil{Harvard-Smithsonian Center for Astrophysics, 60 Garden Street, Cambridge MA 02138; \\ tdimatteo@cfa.harvard.edu}

\author{Steven W. Allen}

\affil{Institute of Astronomy, Madingley Road, Cambridge CB3 OHA, UK; \\ 
swa@ast.cam.ac.uk}

\altaffiltext{1}{{\em Chandra\/} Fellow}

\begin{abstract}

We explore the implications of the discovery of hard, power-law X-ray
sources in the spectra of nearby elliptical galaxies for the origin of
the X-ray background. The spectra of these sources are
consistent with models of thermal bremsstrahlung emission from low
radiative efficiency accretion flows around central supermassive black
holes and are unique in that they approximately match that of the hard
XRB. If such sources, with luminosities consistent with those observed
in nearby ellipticals, are present in most early-type galaxies, then
their integrated emission may contribute significantly to the XRB. These
sources may also contribute to the hard source counts detected in deep
X-ray surveys.

\end{abstract}


\section{Introduction}

Although it is now clear that the Cosmic X-ray Background (XRB)
results from the integrated X-ray emission from many discrete sources,
and that a large fraction of the soft ($0.5-2$ \keV) XRB is produced
by Active Galactic Nuclei (AGN; e.g.,~Hasinger et
al.\markcite{Hetal98}~1998; Schmidt et al.\markcite{Setal98}~1998),
the nature of the sources producing the energetically dominant, hard
($2-60$ \keV) XRB remains largely unknown. Most current models for the XRB 
attempt to explain its origin within the context of AGN unification
schemes and suggest that the XRB arises from the integrated emission of AGN 
with a range of intrinsic absorbing column densities (e.g., Setti \&
Woltjer~\markcite{SW89}~1989; Comastri et al.\markcite{Coetal95}~1995
and references therein). A population of sources with hard
X-ray spectra in the $2-10\keV$ band have been discovered by ASCA and
Beppo-SAX (Boyle et al.~\markcite{Betal98}~1998; Ueda et
al.~\markcite{Uetal98}~1998; Giommi et al.~1998), with a fraction of
these sources showing evidence for heavy obscuration (Fiore et
al.~\markcite{Fietal99}~1999). However, it remains unclear whether
obscured AGN can fully account for the hard XRB.  The most recent
synthesis models, which include the latest constraints on
the luminosity function and evolution of AGN (Miyaji, Hasinger \&
Schmidt 1999), cannot easily reproduce the hard counts observed
in the ASCA (2-10 \keV) and BeppoSAX (5-10 \keV) bands and require a
number ratio of type-2/type-1 AGN much higher than the locally
observed value (Gilli, Risaliti \& Salvati~\markcite{GRS99}~1999).

The discrepancies within the context of AGN synthesis models for 
the XRB suggest the need for an additional population of hard-spectrum
sources. In this {\em Letter}, we explore the implications of the discovery 
of hard, power-law X-ray components in the ASCA spectra of six nearby, 
giant elliptical galaxies (Allen, Di Matteo
\& Fabian~1999; hereafter ADF99). If most early-type galaxies contain a
hard, power-law source with a luminosity of $\sim 10^{40}-10^{42}
\ergps$, as may be extrapolated from the detections of such
components in both active (e.g. M87) and quiescent galaxies, then the 
integrated emission from these sources, distributed over a larger redshift 
interval, can make a significant contribution to the hard XRB.

Dynamical studies of elliptical galaxies indicate the presence of
supermassive black holes in their nuclei, with masses in the range
$10^8-10^{10}$\Msun~(e.g., Magorrian et
al.~\markcite{Metal98}~1998). As discussed by ADF99 and Di Matteo et
al.~(1999a; hereafter DM99), the hard X-ray components
detected in nearby giant ellipticals are likely to be due, at least in
part, to accretion onto their central black holes. In the cores of 
elliptical galaxies,
accretion from the hot interstellar medium may proceed directly into a
hot, low radiative efficiency regime (e.g.,~Fabian \&
Rees\markcite{FR95}~1995). DM99 show that the hard X-ray
components observed in these systems are consistent with 
models of thermal bremsstrahlung
emission from hot, radiatively-inefficient accretion flows, with
temperatures of $50-100\keV$. Given that the XRB spectrum in the
$3-60$ keV band is also well described by a bremsstrahlung spectrum
with $kT \sim 40
\keV$, the hard X-ray sources in elliptical galaxies represent a
unique class of object with emission spectra that closely match that
of the XRB (see also Di Matteo \& Fabian~\markcite{DF97}~1997). 

\section{HARD X-RAY EMISSION FROM ELLIPTICAL GALAXIES}

\subsection{The observed power-law components}

ADF99 discuss ASCA observations of six nearby, giant elliptical
galaxies: M87, NGC 4696 and NGC 1399 (the dominant galaxies of the
Virgo, Centaurus and Fornax clusters) and three other giant
ellipticals in the Virgo Cluster (NGC 4472, NGC 4636 and NGC
4649). All of these galaxies (with the exception of NGC 4696, which
had not previously been studied in as much detail as the other
systems) exhibit clear stellar and gas dynamical evidence for central,
supermassive black holes in their nuclei, with masses in the range
$10^8-10^{10}$\Msun~(e.g., Magorrian et al.~1998). The ASCA spectra
for these systems reveal the presence of hard, power-law emission
components, with energy indices $-0.5<\alpha<0.5$ (weighted-mean value
$\alpha=0.22$; a discussion detailing the reasons why the sources are
unlikely to be heavily obscured AGN is given by ADF99) and intrinsic
$1-10$ keV luminosities of $2 \times 10^{40} - 2
\times 10^{42}$ \ergps. These spectral slopes are harder and the
luminosities are lower than typical values for Seyfert galaxies,
identifying these objects as, potentially, a new class of X-ray
source. The presence of hard components in all six galaxies studied
also suggests that such sources may be ubiquitous in early-type 
galaxies. 

\subsection{An empirical comparison}

We first compare the spectra of the power-law sources observed in nearby
ellipticals with the $1-10 \keV$ XRB. ASCA and BeppoSAX observations 
of the cosmic XRB in the $1-7 \keV$ band can be 
well-described by a simple power-law model with an energy index,
$\alpha = 0.38-0.47$ and a normalization, $I = 8 - 11$ keV
s$^{-1}$cm$^{-2}$sr$^{-1}$ keV$^{-1}$ at 1 keV (Gendreau et
al.~\markcite{Getal95}~1995; Chen, Fabian \&
Gendreau\markcite{Chetal97}~1997; Miyaji et
al.\markcite{MIetal98}~1998; Parmar et al.\markcite{Petal99}~1999).
We have simulated the spectrum obtained by adding a $70-80$ per cent 
contribution to the $1-10$ keV flux from power-law sources with an 
energy index $\alpha=0.22$ (modeling the emission from the elliptical 
galaxies) to the established 
$\approxlt 30$ per cent contribution, in the same band, from unabsorbed 
Seyfert-1 galaxies and QSOs (as determined from ROSAT and ASCA; e.g. 
Schmidt et al.~1998; Boyle et al.~1998). These latter sources have been 
characterized by a power-law spectrum with an intrinsic energy index, 
$\alpha = 0.9$, with a reflection component accounting for 
emission reprocessed by cold material close to the central X-ray sources
(Magdziarz \& Zdziarski 1995). Note that the use of a simpler 
power-law parameterization for the type-1 AGN, with an apparent energy 
index, $\alpha = 0.7$ (Turner \& Pounds\markcite{TP89} 1989) leads to 
similar results.

The simulated spectrum, as would be observed with the ASCA
Solid-state Imaging Spectrometers (SIS) from a 250ks exposure
(matching the total exposure time analyzed by Gendreau et
al.~1995) is shown in Figure 1. The flux in the simulated spectrum 
matches the cosmic XRB flux observed by Gendreau et
al.~(1995). We do not account for the internal background in the SIS
detectors, which provides an additional contribution to the 
total count rate detected by those authors.

Following standard X-ray analysis methods, we have fit the simulated
spectrum in the $1-10 \keV$ range with a simple power-law
model. We find that this model provides a good description of the 
simulated data (reduced $\chi^2 \sim 0.9$ for 200 degrees of 
freedom, after regrouping to a minimum of 20 counts per channel) and 
returns a best-fitting slope of $\alpha=0.40\pm 0.02$ or $\alpha=0.48\pm 
0.02$ (90 per cent errors determined from monte-carlo simulations) for
simulations with 20 and 30 per cent contributions to the $2-10$ keV flux from 
type-1 AGN, respectively. These results are in excellent agreement with those
for the real XRB.

This simple exercise illustrates that, independently of the model used
to explain the hard, power-law emission components in the elliptical 
galaxies, their observed $2-10$ keV spectra match that of the XRB in this 
band, once the expected contributions from QSOs and AGN are also accounted for.

\vspace*{0.5cm}
\vbox{
\label{figure1}
\centerline{\psfig{figure=fig1_tiz.ps,width=9.0truecm,angle=270}}}
\figcaption[]{\footnotesize Simulated ASCA XRB spectrum. The data from
all eight SIS chips have been co-added and a Galactic column density of
$3\times 10^{20}$ atom cm$^{-2}$ is assumed. A 20 per cent contribution
to the $2-10$ keV flux from type-1 AGN is included. The upper panel shows
the simulated data and best fitting power-law model, which has an energy
index of $0.40 \pm 0.02$. The lower panel shows the residuals to the fit in
units of $\chi$. }
\vspace*{0.1cm}

\section{BREMSSTRAHLUNG EMISSION FROM ELLIPTICAL GALAXY NUCLEI}

In previous papers (DM99; Di Matteo et al.~1999b) we have shown that 
the broad-band spectral energy distributions for the nuclear regions of 
the six elliptical galaxies studied by ADF99 can be explained by 
low radiative efficiency accretion models (i.e. advection 
dominated accretion flows or ADAFs; see 
Narayan, Mahadevan \& Quataert \markcite{NMQ98}~1998 and references therein) 
in which accretion occurs from the hot, gaseous halos of the galaxies at 
rates comparable to their Bondi accretion rates, and in which a significant 
fraction of the mass, angular momentum and energy
in the accretion flows is removed by winds (Blandford \& Begelman 
\markcite{BB99}~1999). Within the context of these models, the
systematically hard, observed X-ray spectra can be accounted for by the
energetically-dominant bremsstrahlung emission produced by such flows,
with electron temperatures of $\sim 50-100 \keV$. In this Section, we 
explore the implications for the XRB of this interpretation for the origin 
of the hard X-ray emission in elliptical galaxies.

\subsection{The XRB model}

We consider the integrated emission from unresolved sources, with hard
bremsstrahlung spectra and luminosities consistent with those observed
in the six nearby ellipticals studied by ADF99. We first
constructed a standard coadded source spectrum by combining the results from
fits to the observed spectral energy distributions for the
galaxies with two-temperature ADAF models (including the effects of
winds), as discussed by DM99. The contributions to the co-added
spectrum from the three central cluster galaxies were down-weighted by a
factor $10^2$ to reflect their lower space density. The resulting
co-added source spectrum has a bremsstrahlung 
luminosity of $8 \times 10^{40} \ergps$ (see also Figure 2 in
DM99). This standard source spectrum was then folded with the
appropriate cosmological model to determine the possible, integrated
contribution from such sources to the $2-60\keV$ XRB. (This model is
essentially the same as that described by Di Matteo et
al.\markcite{DM99c}~(1999c) and Di Matteo \& Fabian (1997), but now
including the constraints on the source spectra from DM99; see
Section 5 and Fig.~2 of that paper.)

We assume that the sources are distributed over a redshift range
$z=z_0$ to $z=z_{\rm max}$, and write the comoving spectral emissivity
from such objects as the product $j[E,z]=n(z)L_{\rm E}(z)$, where
$n(z)$ is the comoving number density of X-ray sources, and $L_{\rm
E}(z)$ is the specific luminosity of the individual sources. We adopt
a simple prescription for the redshift evolution of the comoving
emissivity, $j(E,z)=j_0 (E) (1+z)^{k}$, where $j_0 (E)$ is the
model spectrum bremsstrahlung emissivity and $k$ is the evolution
parameter.
The total flux from such objects is then

\begin{equation}
I(E)=\frac{c}{4\pi H_0}\times \int_{z_0}^{z_{\rm max}}\frac{(1+z)^{k-2}}{(1+2q_0z)^{1/2}}j_0[E(1+z)]dz, \nonumber
\end{equation}

where $q_0$ is the deceleration parameter, $H_0$ is the Hubble constant
(we use $q_0=0.5$ and $H_0=50 \kmpspMpc$) and $I(E)$ is the computed
XRB intensity in units of $\keV\ps\psr\pcmsq\keV^{-1}$ at an energy of 1 keV.
For our given source spectrum and fixed values for $z_0$ (we assume $z_0=0$), 
$z_{\rm max}$ and $k$, the only free parameter in Equation 1 is
the local source number density, $n(z_0)$, which we determine by
normalizing $I(E)$ to the observed XRB. 

\vspace*{-1.5cm}
\vbox{
\centerline{\psfig{figure=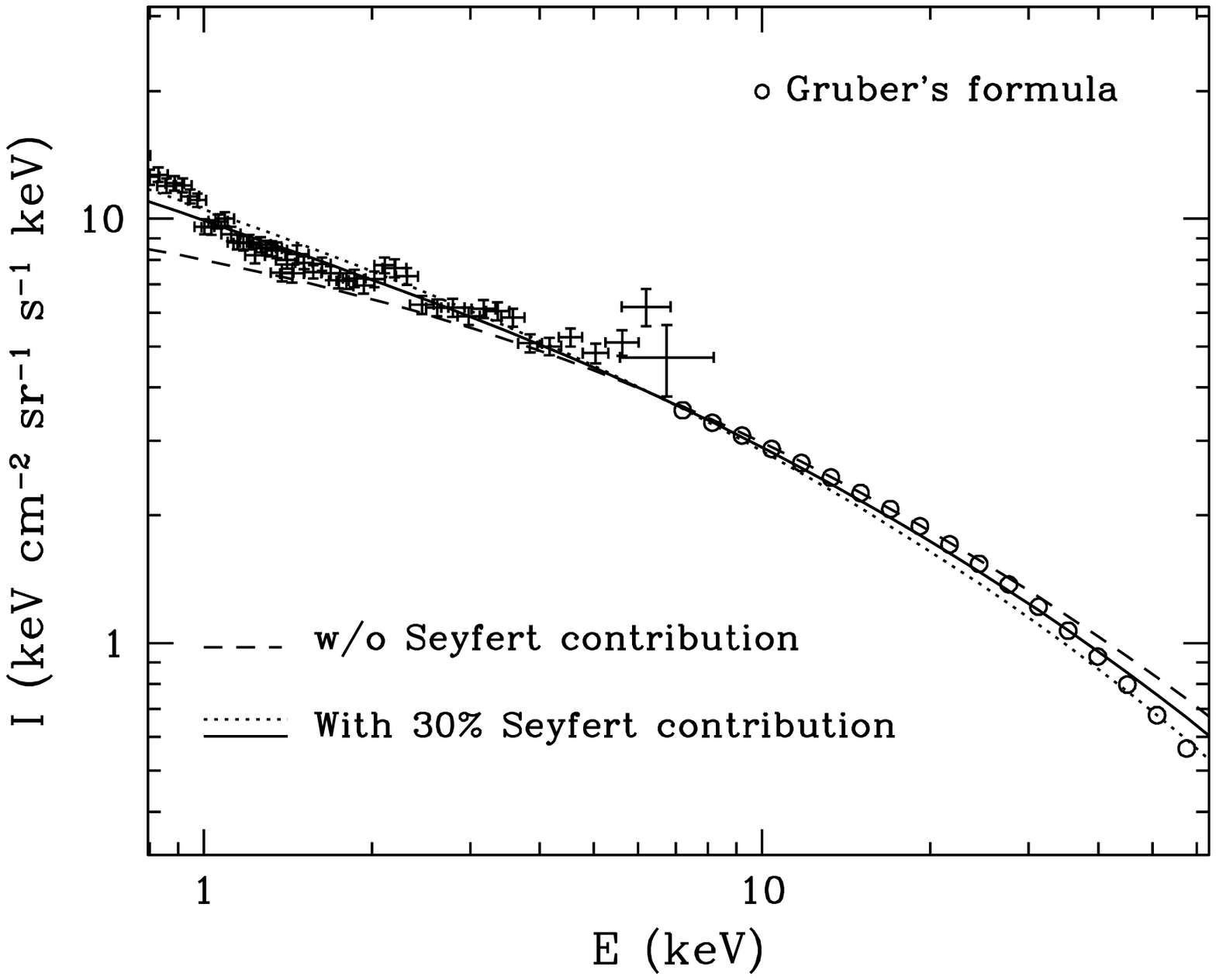,width=10.truecm}}}
\figcaption[]{\footnotesize The XRB models. The dashed curve shows
the result due to bremsstrahlung sources only. The solid and dotted
curves include a 30 per cent contribution to the $2$ keV flux from
type-1 AGN. All curves assume $z_0=0$ and $k=3$. The dashed and solid
curves are for $z_{\rm max} = 1$. The dotted curve is for $z_{\rm max}
= 1.8$. The crosses show the $1-7$ keV XRB spectrum observed with ASCA
(Gendreau et al.~1995; $\alpha = 0.41 \pm 0.03$, $I = 8.9 \pm 0.4$
$\keV\ps\psr\pcmsq\keV^{-1}$ at 1 keV). The positive residual at 6.4
keV is an instrumental emission line). The circles show the Marshall
et al. (1980) result from HEAO A2 data in the $3-60$ \keV~band (a
power-law spectrum with an energy index, $\alpha=0.4$, and an
exponential rollover at $\sim 30$ \keV; see also Gruber 1992). }
\label{xrbfit}
\vspace*{0.1cm}

The XRB spectra predicted by a series of `best fit' models are shown
in Figure~2. The dashed line shows the result due only to
bremsstrahlung sources. The solid line shows the improved result when
we include a 30 per cent contribution to the $2 \keV$ flux from
unabsorbed AGN (corresponding to a $\sim 20$ contribution to the
$2-10$ \keV flux), characterized by a canonical power-law spectrum
with an energy index, $\alpha=0.7$. (Both curves assume $z_{\rm max} =
1.0$ and $k=3$.) In this case, the normalization of the XRB requires a
local comoving number density of bremsstrahlung sources, $n(z_0=0)
\sim 2 \times 10^{-3}\pmpccu$. The dotted line in Figure~2 shows the
result for $z_{\rm max} = 1.8$ (other parameters the same) in which
case $n(z_0=0) = 8 \times 10^{-4}\pmpccu$ (the most distant sources
contribute most to the total model flux). These values for the
comoving number density of emitting sources are in good agreement with
the observed number density of bright, early-type galaxies in the
nearby Universe ($n \sim 10^{-3}\pmpccu$ e.g., Marinoni et
al.\markcite{Metal99}~1999; Heyl et al.\markcite{heetal97}~1997).

\section{DISCUSSION}

We have shown that our model provides a good match to the observed 
XRB, with a required number density of emitting sources in good agreement 
with the 
observed number density of early-type galaxies in the nearby Universe 
(for $z_{\rm max} \approxgt 1$ and $k\sim 3$). The requirement for most/all 
early-type galaxies to have hard X-ray spectra and luminosities consistent 
with the 
objects studied by ADF99 may be relaxed once more realistic models for 
the XRB are considered, including a fractional contribution from heavily 
absorbed Seyfert-2 nuclei (e.g. Gilli et al.~1999). This will then allow 
for a broader range of luminosities and/or a lower number density for the 
bremsstrahlung sources. (In this case the relative contribution from these 
sources to the $2-10$ keV XRB intensity will also be rescaled to $<$ 70 per 
cent.) The required comoving number density of bremsstrahlung sources is also 
decreased as $z_{\rm max}$ is increased towards $z_{\rm max} \sim 2$.

It is important to note that our sources may also contribute to the hard
number counts detected at faint X-ray fluxes by ASCA and BeppoSAX
(e.g. Ueda et al.~1998; Fiore et al.~1999). Detailed synthesis models for the
XRB, which simulate the integrated emission from AGN with a range of absorbing
column densities, have revealed the need for additional hard spectrum sources
to explain the observed number counts (e.g., Gilli et al.~1999). The $2-10$
and $5-10$ keV fluxes associated with the power-law sources detected in the
nearby ellipticals studied by ADF99 range from $0.6-8.7$ and
$0.3-5.0 \times 10^{-12}$\ergpcmsqps, respectively.
Within the context of our model, a significant fraction of the hard number
counts detected at fluxes, $F_{\rm X, 2-10} \approxlt$ a few 
$10^{-14}$\ergpcmsqps; may then arise from sources at low redshift. 
We note that models
postulating the existence of an as yet unobserved population of heavily 
obscured black holes at redshift, $z\sim 2$, also predict X-ray fluxes too 
faint to account for the observed hard counts (Fabian 1999).

The value of $z_{\rm max}$ in our model is constrained by the
effective temperature of the bremsstrahlung emission in the coadded
source spectra: the coadded spectrum must be redshifted to fit the
30~\keV~rollover observed in the XRB. We note that the high-energy
($\approxgt 50 \keV$) spectrum predicted by our model is not firmly
constrained. The presence of winds/outflows associated with low
radiative efficiency accretion flows around supermassive black holes
inevitably causes the X-ray spectra of such flows to be dominated by
bremsstrahlung emission. (Inverse Compton emission is heavily suppressed
in the presence of outflows for any range of $\mdot$; this, with
$\mdot \sim \mdot_{\rm crit}$ for the component sources, allows for
higher individual bremsstrahlung source luminosities than 
considered in the earlier work of Di Matteo \& Fabian.) However, the wind
characterization currently employed in the accretion models is very
basic and the spectra produced cannot be used to perform reliable
statistical fits to the data.  In particular, the presence of an
outflow will significantly affect the density and temperature profiles
in the central regions of an ADAF (c.f. DM99; Quataert \& Narayan
\markcite{QN99}~1999), where the higher energy ($h\nu \approxgt kT$)
emission originates (although the emission in the $2-10
\keV$ band, where the models provide a good match to the power-law
components observed in nearby ellipticals, is virtually unaffected).
Due to these uncertainties the value of $z_{\rm max}$ cannot be tightly
constrained.

The black holes at the centers of nearby elliptical galaxies have masses 
consistent with being 
the remnants of an earlier quasar phase. If, as our work may suggest, these 
systems accrete via low-radiative efficiency accretion flows, then a fraction 
of the hard XRB could be produced after the main quasar phase in galaxies 
and be associated with a change in the dominant accretion mechanism in their
nuclei (see also Di Matteo \& Fabian 1997).  It is interesting, in
this context, that studies with the Hubble Space Telescope have shown
the underlying hosts of essentially all classes of QSOs appear to be
massive, elliptical galaxies (McLure et al.\markcite{M99}~1999).

We stress that our analysis is not intended to explore the full range
of parameter space available to either the accretion flow or XRB
models, or to provide detailed, quantitative results. At present, the
theoretical and observational uncertainties involved in such
calculations are too large to merit such work. The observational 
constraints will, however, be significantly improved in the near
future with data from the Chandra Observatory, XMM and ASTRO-E

\section{CONCLUSIONS}

We have examined the potential importance of the hard, power-law emission 
components detected in the
X-ray spectra of nearby ellipticals for the origin of the hard
XRB.  In previous papers (ADF99 and DM99) we have shown
that these components are likely to be associated with accretion onto
the central, supermassive black holes in the galaxies.  The emission
spectra from these sources can be well-explained by bremsstrahlung
models, with typical temperatures of $50-100 \keV$, resulting from low
radiative-efficiency accretion flows with strong winds. In this paper
we have shown that the application of such emission models to a
plausible redshift distribution of sources, with individual source
luminosities in agreement with the ASCA results for nearby
ellipticals, can account for a significant fraction of the XRB in the
$1-60 \keV$ range, with an implied number density of sources in good
agreement with the observed local number density of early-type
galaxies. We have argued that the emission from these sources may also 
contribute to the hard number counts detected at faint X-ray fluxes 
with ASCA and BeppoSAX.

\acknowledgments 
We thank K. Gendreau for supplying the ASCA spectrum of the XRB and
D. Psaltis for helpful discussions. T.\,D.\,M.\ acknowledges support 
provided by NASA through Chandra Postdoctoral
Fellowship grant number PF8-10005 awarded by the Chandra Science
Center, which is operated by the Smithsonian Astrophysical Observatory
for NASA under contract NAS8-39073.

\end{document}